\documentclass[10pt,aps,prl,twocolumn,showpacs,amssymb,floatfix]{revtex4-1}
\usepackage{amsmath,graphics,epsfig,mathrsfs}
\usepackage{epstopdf}
\usepackage{bm}
\usepackage{color}
\usepackage{soul}
\usepackage{hyperref}
\hypersetup{backref=true,pagebackref=true,hyperindex=true,colorlinks=true,breaklinks=true,urlcolor=
red,linkcolor=blue,bookmarks=true,bookmarksopen=true,citecolor=red}

\begin{document}
\title{Spin Swapping Transport and Torques in Ultrathin Magnetic Bilayers}
\author{Hamed Ben Mohamed Saidaoui$^1$}
\author{A. Manchon$^{1}$}\email{aurelien.manchon@kaust.edu.sa}
\affiliation{$^1$Physical Science and Engineering Division, King Abdullah University of Science and Technology (KAUST), Thuwal 23955-6900, Kingdom of Saudi Arabia}

\begin{abstract}
Planar spin transport in disordered ultrathin magnetic bilayers comprising a ferromagnet and a normal metal (typically used for spin pumping, spin Seebeck and spin-orbit torque experiments) is investigated theoretically. Using a tight-binding model that treats extrinsic spin Hall effect, spin swapping and spin relaxation on equal footing, we show that the nature of spin-orbit coupled transport dramatically depends on ratio between the layers thickness $d$ and the mean free path $\lambda$. While spin Hall effect dominates in the diffusive limit ($d\gg\lambda$), spin swapping dominates in Knudsen regime ($d\lesssim\lambda$). A remarkable consequence is that the symmetry of the spin-orbit torque exerted on the ferromagnet is entirely different in these two regimes.
\end{abstract}
\maketitle
\paragraph{Introduction}

Spin-orbit coupling is responsible for a wide variety of phenomena that have attracted a huge amount of attention recently \cite{Jungwirth2012,Manchon2015}. Among the most prominent phenomena, one can mention the inverse spin galvanic effect, i.e. the electrical generation of a nonequilibrium magnetization \cite{Ivchenko1978} and the spin Hall effect \cite{Dyakonov1971,Vignale2010}, i.e. the conversion of an unpolarized charge current into a pure spin current. The nature spin Hall effect has been scrutinized intensively lately due to its central role in spintronics. While the original theory was based on carrier scattering against extrinsic spin-orbit coupled impurities \cite{Dyakonov1971}, the importance of the band structure's Berry curvature has been recently unveiled, producing large dissipationless (i.e. scattering independent) spin Hall effects \cite{Sinova2004}. Although the proper theoretical treatment of this effect in the diffusive limit continues raising debates \cite{Ado2015}, experiments tend to confirm the importance of intrinsic spin Hall effect in 4$d$ and 5$d$ transition metals \cite{Vila2007,Du2014}. In the opposite limit, i.e. when the system size becomes comparable to the mean free path (so-called Knudsen regime), the nature of spin Hall effect changes subtly as quantum and semiclassical size effects emerge \cite{Wang2014a}. An accurate description of spin-orbit coupled transport in this regime is crucial as ultrathin normal metal/ferromagnet bilayers (e.g. Pt/NiFe etc.) are now commonly used in spin pumping \cite{Saitoh2006}, spin Seebeck effect \cite{Uchida2010} and spin-orbit torque \cite{Miron2011,Liu2012} measurements.

\begin{figure}[h!]
  \includegraphics[width=0.48\textwidth]{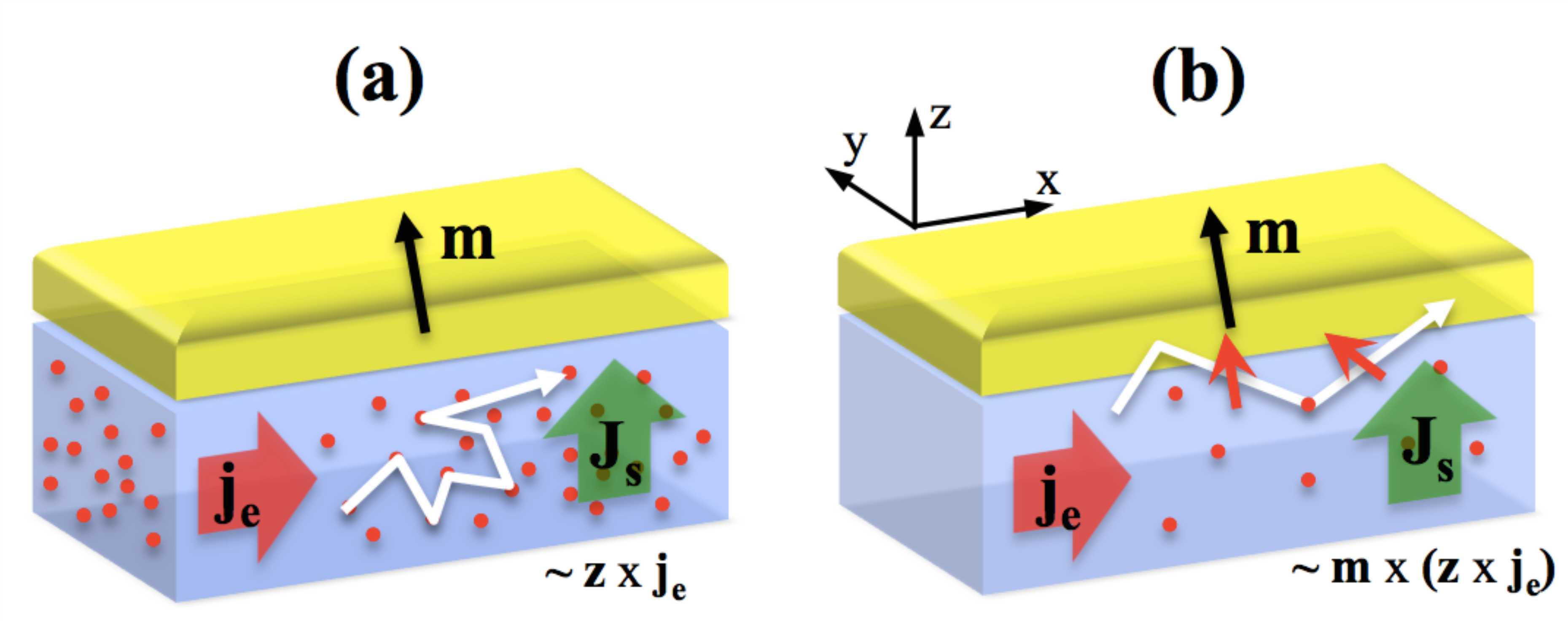}
	\caption{(Color online) Schematics of (a) spin Hall and (b) spin swapping effects in a bilayer composed of a normal metal (blue) and a ferromagnet (yellow) with magnetization ${\bf m}$ in the diffusive and Knudsen regimes, respectively. The charge current ${\bf j}_e$ is injected in the plane of the layers and results in a spin current ${\bf J}_s$ flowing perpendicular to the interface.\label{Fig1}}
\end{figure}

The limitations of the current models of spin Hall effect are best illustrated by the puzzles raised by spin-orbit torque experiments in such ultrathin multilayers. In magnetic systems lacking inversion symmetry, spin-orbit coupling enables the electrical control of the magnetic order parameter \cite{Bernevig2005,Manchon2008,Zelezny2014,Brataas2014}. This spin-orbit mediated torque has been observed in a various materials combinations involving heavy metals \cite{Miron2011,Liu2012}, oxides \cite{Qiu2015} and topological insulators \cite{Mellnik2014}. Experimentally, the torque possesses two components referred to as damping torque (even in magnetization direction) and field-like torque (odd in magnetization direction). Consensually, the damping torque is associated with the spin Hall effect occurring in the bulk of the heavy metal \cite{Haney2013} [see Fig. \ref{Fig1}(a)], while the field-like torque arises from the inverse spin galvanic effect induced by spin-orbit coupling present at the interface between the heavy metal and the ferromagnet \cite{Bernevig2005,Manchon2008}. However, experiments have reported complex material dependence parameters that are not accounted for by these models \cite{Avci2014,Kim2012}. In particular, sizable field-like torques have been observed in systems where interfacial spin-orbit coupling is expected to be small \cite{Pai2014}. In spite of intense theoretical efforts \cite{Freimuth2014,Li2015}, no efficient mechanisms related to spin Hall effect and able to generate sizable field-like torque have been identified \cite{Haney2013}.

In this Letter, we theoretically demonstrate that the nature of extrinsic spin-orbit coupled transport in disordered ultrathin magnetic bilayers dramatically depends on the transport regime. When disorder is strong and transport is diffusive, spin Hall effect dominates leading to a damping torque in agreement with the widely accepted physical picture \cite{Haney2013}. In contrast, when disorder is weak and the system size is of the order of the carrier mean free path, spin swapping \cite{Lifshits2009} becomes increasingly important, leading to a dominant field-like torque.

\paragraph{General principles}
Consider a metallic bilayer composed of a spin-orbit coupled normal metal and a ferromagnet without spin-orbit coupling (see Fig. \ref{Fig1}). A current ${\bf j}_e$ is injected in the plane of the bilayer and exert a torque on the ferromagnet. Disregarding interfacial spin-orbit coupling, two spin-orbit coupled transport phenomena are present in this system. First, due to intrinsic and/or extrinsic spin Hall effect in the normal metal, a spin current flows along the normal to the interface ${\bf z}$ with a spin polarization along $({\bf z}\times{\bf j}_e)$. This results in a spin torque on the form $\sim{\bf m}\times[({\bf z}\times{\bf j}_e)\times{\bf m}]$, which is of {\em damping-like} form \cite{Haney2013} (even in magnetization), see Fig. \ref{Fig1}(a). In addition, electrons flowing in the ferromagnet acquire a spin polarization along ${\bf m}$ and may scatter towards the normal metal. Once in the normal metal, these electrons experience spin swapping: upon scattering on spin-orbit coupled impurities, they experience a spin-orbit field oriented normal to the scattering plane [i.e. along $({\bf z}\times{\bf j}_e)$] and about which their spin precess \cite{Lifshits2009,Saidaoui2015}. Upon this reorientation, a spin current polarized along ${\bf m}\times({\bf z}\times{\bf j}_e)$ is injected into the ferromagnet and induces a {\em field-like} torque (odd in magnetization), see Fig. \ref{Fig1}(b), even in the absence of interfacial spin galvanic effect. Since these two effects operate in distinct disorder regimes, namely spin Hall effect necessitates strong disorder while spin swapping survives even for weak disorder \cite{Saidaoui2015}, the nature of the torque should dramatically change from one regime to the other.\par

\paragraph{Numerical results}
To investigate these effects quantitatively, we computed the spin transport in a magnetic bilayer using a tight-binding model \cite{kwant,Saidaoui2015}. The system is a two-dimensional square lattice connected laterally to external leads. The full Hamiltonian of the central system reads
\begin{eqnarray}
{\hat H}& = &\sum_{i,j,\sigma,\sigma'}\{(\epsilon_{ij}\delta_{\sigma\sigma'}+\frac{\Delta_{ij}}{2}{\bf m}\cdot{\hat{\bm\sigma}}_{\sigma\sigma'}){\hat c}_{i,j,\sigma}^+ {\hat c}_{i,j,\sigma'}+h.c.\}\nonumber\\
&&-\sum_{i,j,\sigma}t_{\rm N}({\hat c}_{i+1,j,\sigma}^+ {\hat c}_{i,j,\sigma}+{\hat c}_{i,j+1,\sigma}^+ {\hat c}_{i,j,\sigma}+h.c.)\nonumber\\
&&-\sum_{i,j}t^{i-1,j}_{i,j-1}({\hat c}_{i,j,\uparrow}^+{\hat c}_{i-1,j-1,\uparrow}-{\hat c}_{i,j,\downarrow}^+{\hat c}_{i-1,j-1,\downarrow})\nonumber\\
&&-\sum_{i,j}t^{i,j}_{i-1,j-1}({\hat c}_{i,j-1,\uparrow}^+{\hat c}_{i-1,j,\uparrow}-{\hat c}_{i,j-1,\downarrow}^+{\hat c}_{i-1,j,\downarrow}).\label{eq:H}
\end{eqnarray}
Here the first term at the right-hand side of Eq. (\ref{eq:H}) is the spin-independent onsite energy in which $\epsilon_{ij}=\epsilon_0+\gamma_{ij}$, $\epsilon_0$ being the onsite energy and $\gamma_{ij}\in[-\Gamma/2,\Gamma/2]$ a random potential of strength $\Gamma$ that introduces disorder in the system. The second term is the exchange interaction ($\equiv\Delta_{ij}$) between the spin of the carriers and the local magnetic moment of direction ${\bf m}$ on site $(i,j)$. The third term in the Hamiltonian corresponds to the nearest neighbor hopping energy ($\equiv t_{\rm N}$). The last two terms are the next-nearest neighbor hopping parameters that account for the disorder-driven spin-orbit coupled scattering. The next-nearest neighbor hopping parameter reads $t^{i,j}_{i',j'}=it_{\rm N}\alpha(\epsilon_{i,j}-\epsilon_{i',j'})$, where $\alpha$ is the dimensionless spin-orbit coupling strength. The operator ${\hat c}_{i,j,\sigma}^+$ (${\hat c}_{i,j,\sigma}$) creates (annihilates) a particle with spin $\sigma$ at position $(i,j)$. This approach models spin Hall effect, spin swapping and spin relaxation on equal footing \cite{Saidaoui2015} and one can tune the relative strength between spin Hall effect and spin swapping by changing the disorder strength $\Gamma$ (and thereby the mean free path) and the spin-orbit coupling strength $\alpha$.\par

\begin{figure}[h!]
  \includegraphics[width=0.55\textwidth]{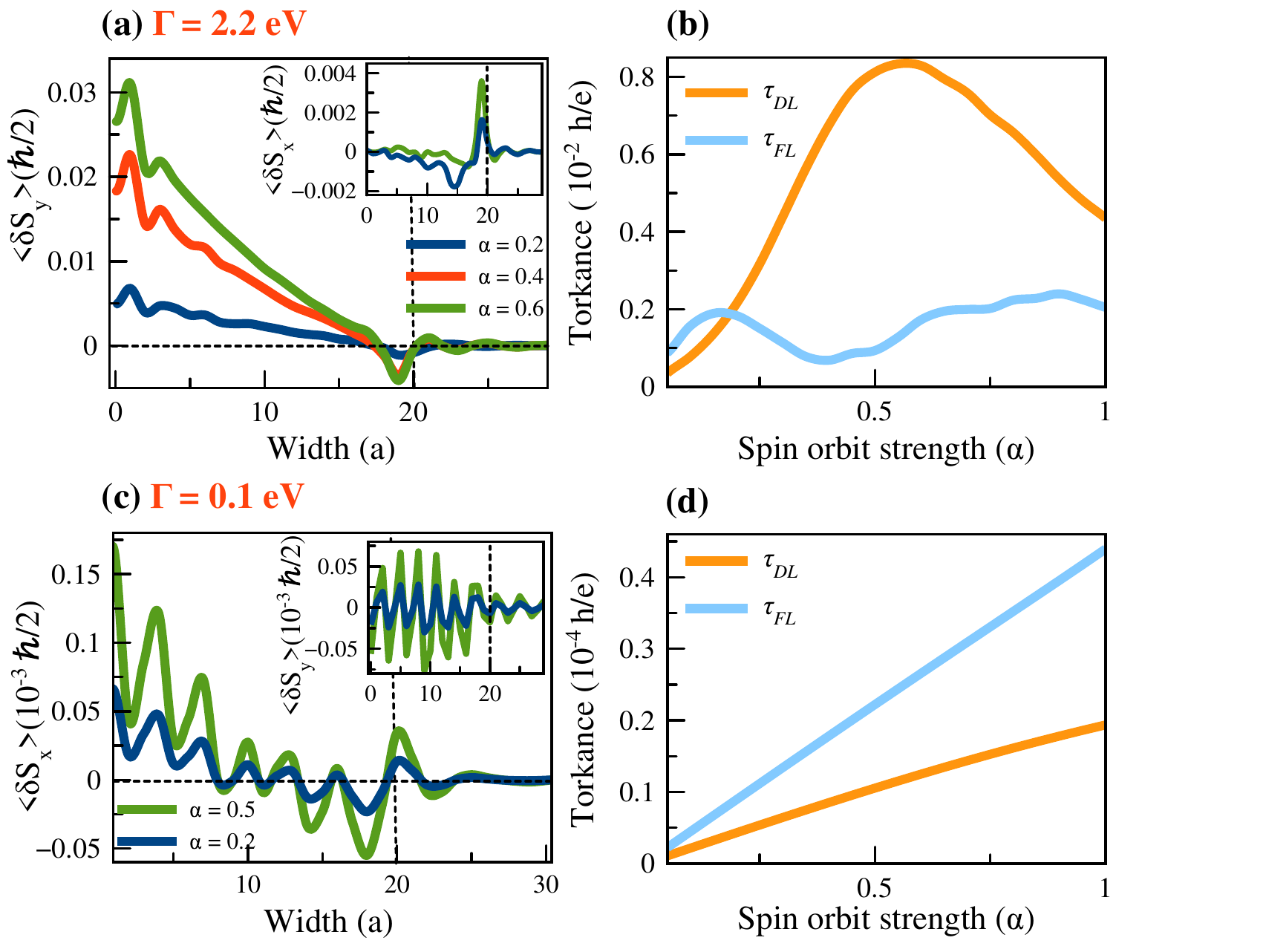}
	\caption{(Color online) (a,c) Spin density profile along the magnetic bilayer width for strong ($\Gamma$=2.2 eV) and weak disorder ($\Gamma=0.1$ eV) regimes. The vertical dashed line separates the normal metal (left) from the ferromagnetic layer (right). The main panels (inset) represent the largest (smallest) spin density component for various $\alpha$. (b,d) Corresponding spin torkance components as a function of $\alpha$. The parameters are $t_{\rm N}=\epsilon_0=\Delta=1$ eV and ${\bf m}={\bf z}$.}
	\label{Fig2}
\end{figure}

\begin{figure}[t]
  \includegraphics[width=0.4\textwidth]{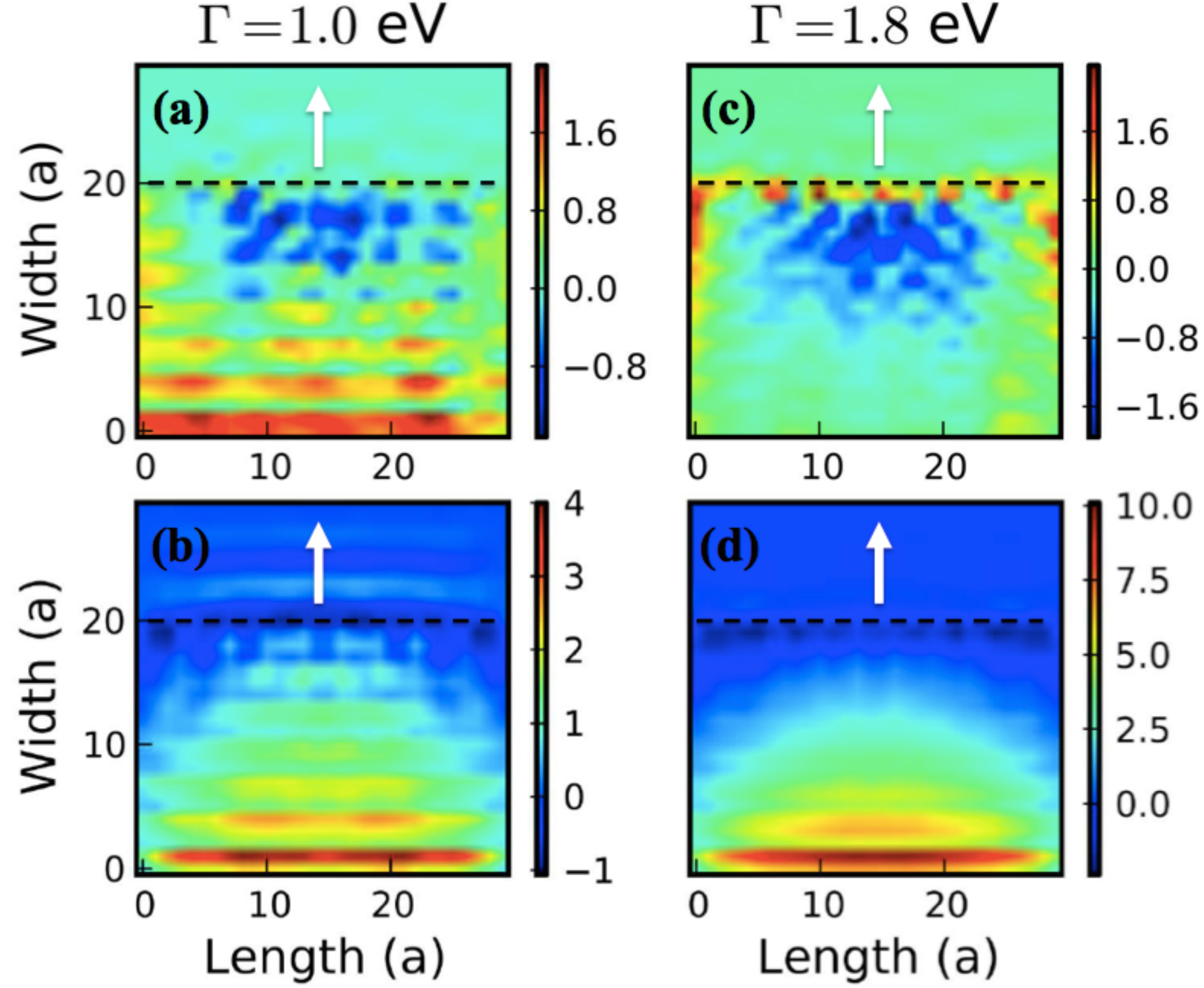}
	\caption{(Color online) Two-dimensional mapping of the spin density components, (a,c) $\delta S_x$ and (b,d) $\delta S_y$, in (a,b) weak and (c,d) strong disordered regimes. The dashed line represents the interface between the normal metal and the ferromagnet and the white arrow indicates the direction of the magnetization. The parameters are the same as in Fig. \ref{Fig2}.}
	\label{Fig3}
\end{figure}

\begin{figure}[h]
  \includegraphics[width=0.4\textwidth]{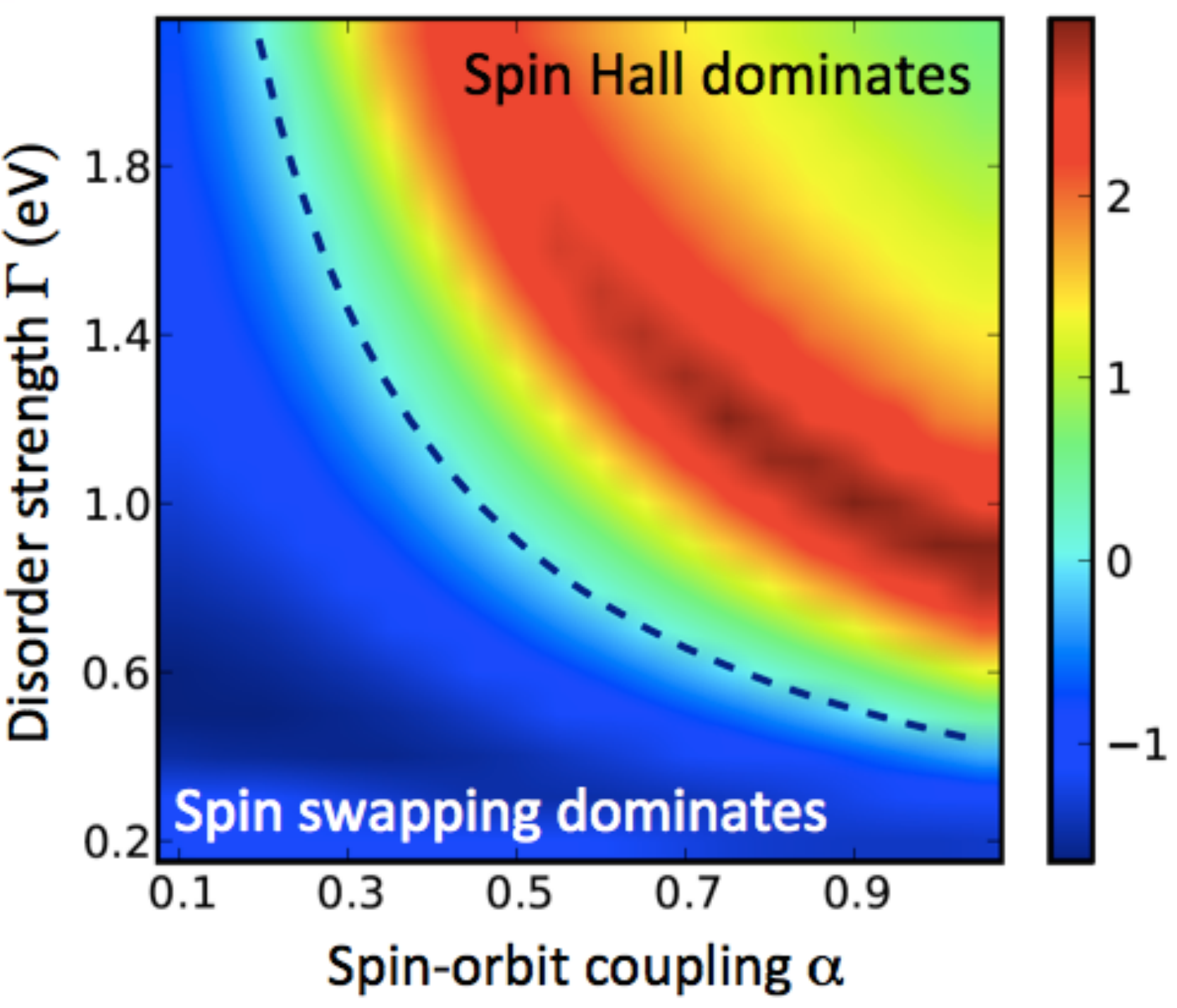}
	\caption{(Color online) Ratio between the magnitude of the field-like torque and damping-like torque, $\tau_{FL}/\tau_{DL}$, as a function of $\Gamma$ and $\alpha$. The ratio is given in logarithmic scale and the dashed line indicates $\tau_{FL}/\tau_{DL}$=1. The parameters are the same as in Fig. \ref{Fig2}.}
	\label{Fig4}
\end{figure}

Let us now consider the current-driven spin density in a two-dimensional bilayer in ({\bf x},{\bf z}) plane (see Fig. \ref{Fig1}). The width of the ferromagnet (normal metal) is 10 $a$ (20 $a$), and the length of the bilayer is 30 $a$, where $a$ is the square lattice parameter. Figure \ref{Fig2}(a) shows the nonequilibrium spin density profile along the bilayer width obtained for a strongly disordered system ($\Gamma=2.2$ eV) and various spin-orbit coupling strengths, $\alpha$. Figure \ref{Fig3}(c,d) displays the corresponding two dimensional map of the spin density components, $\delta S_x$ and $\delta S_y$, respectively. The spin density is mainly aligned along $\delta S_y$ (main panel) and has a small $\delta S_x$ contribution (inset). Remarkably, $\delta S_y$ smoothly accumulates over the layer width, as expected from spin Hall effect in the diffusive regime. Notice though that oscillations stemming from quantum coherence survive even for this amount of disorder, as no extrinsic quantum dephasing is introduced. The small $\delta S_x$ component is confined at the interface, which illustrates its spin swapping origin: it only survives within a distance of the order of the mean free path, as illustrated in Fig. \ref{Fig1}(b). As a consequence, in diffusive regime, the efficiency of the torque (torkance) exerted on the magnetic layer ${\bm \tau}/G=\int d\Omega \delta{\bf S}\times{\bf m}$ ($\Omega$ is the volume of the magnet, $G$ is the conductance of the bilayer) is dominated by a damping-like component $\tau_{\rm DL}$, i.e. $\sim {\bf m}\times[({\bf z}\times{\bf j}_e)\times {\bf m}]$. Nonetheless, when reducing the spin-orbit strength the spin Hall effect decreases and one observes a transition between damping-dominated torque ($\tau_{\rm DL}$) to field-like-dominated torque ($\tau_{\rm FL}$), as illustrated in Fig. \ref{Fig2}(b). This crossover occurs because spin relaxation, which is detrimental to spin swapping \cite{Saidaoui2015}, decreases with $\alpha$ thereby enhancing spin swapping. This region can be widened by decreasing the disorder strength, as shown in Fig. \ref{Fig4}.

The case of weak disorder is even more remarkable, as shown in Fig. \ref{Fig2}(c) [two dimensional mapping is given in Fig. \ref{Fig3}(a,b)]. Phase coherence results in quantum oscillations of both $\delta S_x$ (main panel) and $\delta S_y$ (inset). Nevertheless, while the oscillations of $\delta S_y$ in the normal metal are symmetric with respect to the center of the layer (a reminiscence of the standing nature of the wave functions), thereby resulting in a vanishing spin current injection, the oscillations of $\delta S_x$ are distorted and result in an effective spin current injection into the adjacent ferromagnet. As a consequence, the torkance is dominated by the field-like component, $\tau_{\rm FL}$, i.e. $\sim {\bf m}\times({\bf z}\times{\bf j}_e)$ [right panel of Fig. \ref{Fig2}(d)] for all $\alpha$, in agreement with the phenomenological discussion provided above.

In Fig. \ref{Fig4}, the ratio $\tau_{FL}/\tau_{DL}$ is displayed as a function of disorder and spin-orbit coupling strengths. Interestingly, we find that the torque is dominated by the field-like component in the weak disorder/weak spin-orbit coupling regime, while it is dominated by the damping like component in the strong disorder and/or strong spin-orbit coupling regime. These different phases can be directly attributed to the spatial dependences shown in Fig. \ref{Fig2}. When spin Hall effect dominates (diffusive regime), the torque is mostly damping-like, and when spin swapping dominates (Knudsen regime), the torque is mostly field-like. This behavior has been reproduced by varying the thickness of the normal metal while keeping the disorder fixed (not shown). These simulations demonstrate that in ultrathin bilayers field-like torques do not necessarily arise from interfacial spin galvanic effect, but can emerge due to spin dependent scattering in the normal metal. A necessary condition is that the thickness of the normal metal ought to be of the order of the mean free path.

\paragraph{Drift-diffusion model} Let us now address the nature of the spin swapping torque in the diffusive regime. Indeed, it may look surprising that spin swapping dramatically decreases when disorder is strong. The spin-orbit coupled spin transport in the normal metal can be modeled using the spin diffusion equation developed in Ref. \onlinecite{Shchelushkin2005} in the 1$^{st}$ Born approximation
\begin{eqnarray}\label{eq:jc}
&&e{\bf j}_{e}/\sigma_{\rm N}=-{\bm\nabla}\mu_c+\frac{\alpha_{\rm sh}}{2}{\bm \nabla}\times{\bm\mu},\\\label{eq:js}
&&e^2{\cal J}_{\rm s}^i/\sigma_{\rm N}=-{\bm \nabla}\frac{\mu_i}{2}-\alpha_{\rm sh}{\bf e}_i\times{\bm\nabla}\mu_c+\frac{\alpha_{\rm sw}}{2}{\bf e}_i\times({\bm\nabla}\times{\bm \mu}),
\end{eqnarray}
where $\sigma_{\rm N}$ is the bulk conductivity, $\alpha_{\rm sh}=\alpha/\lambda k_{\rm F}$ is the Hall angle from side jump scattering (within 1$^{st}$ Born approximation, skew scattering is absent) and $\alpha_{\rm sw}=2\alpha/3$ is the spin swapping coefficient. $\lambda$ and $k_{\rm F}$ are the mean free path and Fermi wavevector, respectively. $\mu_c$ and ${\bm\mu}$ are the spin-independent and spin-dependent electrochemical potentials, related to the charge and spin accumulation by  $\mu_c=n/e{\cal N}$ and ${\bm\mu}=\delta{\bf S}/e{\cal N}$, and ${\cal N}$ is the density of state. Note that ${\bf j}_{e}$ is the current density vector whereas ${\cal J}_{\rm s}$ is the spin density tensor and ${\cal J}_{\rm s}^i$ is the $i$-th spin component of the spin current. This set of equations is combined with the spin and charge accumulation continuity equations ${\bm \nabla}\cdot{\bf j_c}=0$ and ${\bm \nabla}\cdot{\cal J}_s=-{\bm\mu}/\tau_{\rm sf}$ where $\tau_{\rm sf}$ is the spin relaxation time. The spin transport in the ferromagnet is modeled by similar drift-diffusion equations \cite{Haney2013}. \par

To model the torque exerted on the ferromagnet, we assume that the spin dephasing in the magnetic layer is so short that the incoming spin current is entirely absorbed within a few monolayers from the interface. The boundary conditions are then written \cite{Brataas2006}
\begin{eqnarray}
{\bf j}_e=&&g\Delta\mu_c+\gamma g\Delta\mu_\|,\;\;{\cal J}_{s,z}^\|=\gamma g\Delta\mu_c+ g\Delta\mu_\|,\\
{\cal J}_{s,z}^\bot=&&2(-g_r^{\uparrow\downarrow}\Delta\mu_{\rm op}+g_i^{\uparrow\downarrow}\Delta\mu_{\rm ip}){\bf m}\times{\bf y}\nonumber\\
&&-2(g_r^{\uparrow\downarrow}\Delta\mu_{\rm ip}+g_i^{\uparrow\downarrow}\Delta\mu_{\rm op}){\bf m}\times({\bf y}\times{\bf m}),
\end{eqnarray}
where ${\cal J}_{s,z}^\|={\cal J}_{s,z}\cdot{\bf m}$ and ${\cal J}_{s,z}^\bot={\cal J}_{s,z}-{\cal J}_{s,z}^\|\cdot{\bf m}$ is the spin current transverse to the magnetization ${\bf m}$. We define $g=(g_\uparrow+g_\downarrow)/2$ and $\gamma=(g_\uparrow-g_\downarrow)/2g$, $g_s$ being the interfacial conductance for spin $s$ and $g^{\uparrow\downarrow}=g_r^{\uparrow\downarrow}+ig_i^{\uparrow\downarrow}$ is the (complex) mixing conductance. The algebra to obtain the interfacial spin current is cumbersome but does not present technical difficulties. We find that the torque possesses two contributions, ${\bm \tau}={\bm \tau}_{\rm sh}+{\bm \tau}_{\rm sw}$, associated with spin Hall and spin swapping respectively, and
\begin{eqnarray}
{\bm \tau}_{\rm sh}&=&\frac{\tilde{\alpha}_{\rm sh}j_{\rm N}}{D_\theta}\eta_0\left[(\tilde{g}_r^{\uparrow\downarrow}-|\tilde{g}^{\uparrow\downarrow}|^2){\bf m}\times({\bf y}\times{\bf m})-\tilde{g}_i^{\uparrow\downarrow}{\bf m}\times{\bf y}\right],\nonumber\\
&&\label{eq:tsh}\\
{\bm \tau}_{\rm sw}&=&\alpha_{\rm sw}\frac{\tilde{\alpha}_{\rm sh}j_{\rm N}}{D_\theta}\left[\tilde{g}_r^{\uparrow\downarrow}\eta_{\rm N}m_z{\bf m}\times{\bf x}+\tilde{g}_i^{\uparrow\downarrow}\eta_{\rm N}m_z{\bf m}\times({\bf x}\times{\bf m})\right.\nonumber\\
&&\left.+|\tilde{g}^{\uparrow\downarrow}|^2m_x{\bf m}\times{\bf z}\right],\label{eq:tsw}\\
D_\theta&\approx&\eta_0(1-\tilde{g}_r^{\uparrow\downarrow})^2+\alpha_{\rm sw}(\eta_{\rm N}+\tilde{g}_r^{\uparrow\downarrow})(1-\tilde{g}_r^{\uparrow\downarrow})\sin^2\theta.\nonumber
\end{eqnarray}
In order to keep the notation compact, we defined the effective spin Hall angle $\tilde{\alpha}_{\rm sh}=\alpha_{\rm sh}(1-\cosh^{-1} d_{\rm N}/\lambda^{\rm N}_{\rm sf})$ and normalized mixing conductances $\tilde{g}_j^{\uparrow\downarrow}=4\tilde{\lambda}^{\rm N}_{\rm sf}g_j^{\uparrow\downarrow}/\sigma_{\rm N}$, where $\tilde{\lambda}^{\rm N}_{\rm sf}=\lambda^{\rm N}_{\rm sf}/\tanh (d_{\rm N}/\lambda^{\rm N}_{\rm sf})$ is the effective spin diffusion length of the normal metal. Finally, 
$\eta_{\rm N}=\frac{4\tilde{\lambda}^{\rm N}_{\rm sf}/\sigma_{\rm N}}{4\tilde{\lambda}^{\rm F}_{\rm sf}/\sigma_{\rm F}+1/g}$ and $\eta_0=1+\eta_{\rm N}$. Here $\sigma_{\rm F}$ and $\lambda_{\rm sf}^{\rm F}$ are the conductivity and spin diffusion length of the ferromagnetic layer, and $j_{\rm N}$ is the charge current density flowing in the normal metal.\par

The spin Hall torque, ${\bm \tau}_{\rm sh}$ [Eq. (\ref{eq:tsh})], solely arises from spin Hall effect ($\propto \alpha_{\rm sh}$) and produces the regular damping torque ${\bf m}\times({\bf y}\times{\bf m})$, with a small contribution to the field-like torque ${\bf m}\times{\bf y}$ \cite{Haney2013}. In the presence of spin swapping, these two torques are renormalized by the denominator $D_{\theta}$ that depends on $\sin^2\theta$ through the spin swapping coefficient $\alpha_{\rm sw}$ (where $\cos\theta={\bf m}\cdot{\bf z}$). More interestingly, the spin swapping torque, ${\bm \tau}_{\rm sw}$ [Eq. (\ref{eq:tsw})], arises from the interplay between spin swapping and spin Hall effect ($\propto\alpha_{\rm sw}\alpha_{\rm sh}$) and generates three additional torque components. Assuming $\tilde{g}_i^{\uparrow\downarrow}\ll\tilde{g}_r^{\uparrow\downarrow}\ll1$, to the leading order in $\tilde{g}_r^{\uparrow\downarrow}$, the torques reduce to
\begin{eqnarray}
{\bm \tau}_{\rm sh}&\approx&\frac{\tilde{\alpha}_{\rm sh}j_{\rm N}}{D_\theta}\eta_0\tilde{g}_r^{\uparrow\downarrow}{\bf m}\times({\bf y}\times{\bf m}),\label{eq:tsh2}\\
{\bm \tau}_{\rm sw}&\approx&\alpha_{\rm sw}\frac{\tilde{\alpha}_{\rm sh}j_{\rm N}}{D_\theta}\tilde{g}_r^{\uparrow\downarrow}\eta_{\rm N}m_z{\bf m}\times{\bf x},\label{eq:tsw2}
\end{eqnarray}
and the ratio between these two contributions is given by $\alpha_{\rm sw}\frac{\eta_{\rm N}}{1+\eta_{\rm N}}$. Therefore, in the diffusive regime, since most of the current flows in the bulk of the normal metal the contribution of the spin swapping close to the interface is vanishingly small [as shown in the inset of Fig. \ref{Fig2}(b)] and the only manner spin swapping contributes to the torque is through its interplay with spin Hall effect, which is second order in spin-orbit coupling and reasonably much smaller than the spin Hall effect in agreement with our tight-binding calculations.\par

\paragraph{Discussion and perspectives}

These results are of direct relevance for experiments on current-driven spin-orbit torque \cite{Miron2011,Liu2012}, but also spin pumping \cite{Saitoh2006} and spin Seebeck \cite{Uchida2010} measurements in ultrathin magnetic multilayers. As a matter of fact, most of these investigations are conducted on multilayers comprising metals with thicknesses from 10 nm down to less than 1 nm (see, e.g. Refs. \cite{Avci2014,Kim2012}). In sputtered thin films, the grain size ranges from 5 to 10 nm, which suggests that the transport is not diffusive and that extrinsic spin swapping can lead to sizable field-like torque even in the absence of interfacial inverse spin galvanic effect. In addition, in the numerical calculations reported here the effect of intrinsic (Berry phase-induced) spin Hall effect, dominant in 4$d$ and 5$d$ transition metals, was disregarded. Then, one can reasonably expect that the spin-orbit coupling in the band structure should also induce spin swapping \cite{Sadjina2012}. This effect is well-known in semiconductors where the coherent precession about the local spin-orbit field induces, e.g., D'yakonov-Perel spin relaxation. Intrinsic spin swapping can be estimated using {\em ab initio} calculations but it may be difficult to disentangle this effect from interfacial inverse spin galvanic effect \cite{Freimuth2014}.\par

We conclude this work by commenting on the impact of spin swapping on spin pumping, the Onsager reciprocal of spin transfer torque. When excited by a radio-frequency field (or by thermal magnons), the precessing magnetization pumps a spin current, polarized along $\sim {\bf m}\times\partial_t {\bf m}$, into the normal metal \cite{Tserkovnyak2002}. Such a spin current can be converted into a charge current through inverse spin Hall effect \cite{Saitoh2006}, but it also enhances the magnetic damping of the ferromagnet \cite{Tserkovnyak2002}. Upon spin swapping this pumped spin current is converted into another spin current polarized along ${\bf y}\times({\bf m}\times\partial_t {\bf m})$ (${\bf y}$ being the direction of the spin-orbit field perpendicular to the scattering plane). While this new spin current does not contribute to additional electric signal, it should produce a corrective damping torque on the form $\sim m_y\partial_t{\bf m}$, i.e. an anisotropic magnetic damping. This effect vanishes by symmetry in homogeneous ferromagnets, but is expected to survive in magnetic domain walls resulting in unconventional magnetic damping. Further theoretical investigations and experimental explorations are necessary to uncover the full implications of this effect.

\begin{acknowledgements}
A.M. acknowledges inspiring discussions with T. Valet and H.B.M.S. thanks S. Feki for his valuable technical support. This work was supported by the King Abdullah University of Science and Technology (KAUST).
\end{acknowledgements}

\end{document}